\documentclass[aps,prd,preprint,superscriptaddress,showpacs,floatfix,nofootinbib]{revtex4}

\usepackage{graphicx}

\usepackage{graphicx,epsfig}

\def\bwt{\begin{widetext}}

\def\ewt{\end{widetext}}

\def\be{\begin{equation}}

\def\ee{\end{equation}}

\def\bea{\begin{eqnarray}}

\def\eea{\end{eqnarray}}

\def\bean{\begin{eqnarray*}}

\def\eean{\end{eqnarray*}}

\def\bary{\begin{array}}

\def\eary{\end{array}}

\def\bit{\begin{itemize}}

\def\eit{\end{itemize}}

\def\su5u1{SU(5) \times U(1)}

\def\fsu5u1{SU(5) \times U(1)'}

\def\so10{SO(10)}

\def\sq20{SO(10) \times SO(10)}

\usepackage[centertags]{amsmath}

\usepackage{amssymb}

\newcommand{\Z}{{\mathbb Z}}

\begin{document}

\title{Realistic Type IIB Supersymmetric Minkowski Flux Models 
without the Freed-Witten Anomaly}

\author{Ching-Ming Chen}

\affiliation{George P. and Cynthia W. Mitchell Institute for
Fundamental Physics, Texas A$\&$M University, College Station, TX
77843, USA }

\author{Tianjun Li}

\affiliation{George P. and Cynthia W. Mitchell Institute for
Fundamental Physics, Texas A$\&$M University, College Station, TX
77843, USA }

\affiliation{ Institute of Theoretical Physics, Chinese Academy of
Sciences, Beijing 100080, P. R. China}

\author{Dimitri V. Nanopoulos}

\affiliation{George P. and Cynthia W. Mitchell Institute for
Fundamental Physics, Texas A$\&$M University, College Station, TX
77843, USA }

\affiliation{Astroparticle Physics Group, Houston Advanced
Research Center (HARC), Mitchell Campus, Woodlands, TX 77381, USA}

\def\athens{Academy of Athens, Division of Natural Sciences, 
28 Panepistimiou Avenue, Athens 10679, Greece}

\affiliation{\athens}

%\date{\today}

%%%%%%%%%%%%%%%%%%%%%%%%%%%%%%%%%%%%%%%%%%%%%%%%%%%%%%%%%%%%%%%%%%%%%%%%%%%%

\begin{abstract}

We show that there exist supersymmetric Minkowski 
vacua on Type IIB toroidal orientifold with general 
flux compactifications where the RR tadpole 
cancellation conditions can be relaxed and the 
Freed-Witten anomaly can be cancelled elegantly. 
We present a realistic Pati-Salam like flux model 
without the Freed-Witten anomaly. At the string 
scale, we can break the gauge symmetry down to the 
Standard Model (SM) gauge symmetry, achieve the 
gauge coupling unification naturally, and decouple 
all the extra chiral exotic particles. Thus, 
we have the supersymmetric SMs with/without 
SM singlet(s) below the string scale. Also,
we can explain the SM fermion masses and mixings.
In addition, the unified gauge coupling and the real
parts of the dilaton and K\"ahler moduli are
functions of the four-dimensional dilaton. 
The complex structure moduli and one linear 
combination of the imaginary parts of the K\"ahler 
moduli can be determined as functions of the 
fluxes and the dilaton.

\end{abstract}


\preprint{ACT-07-08, MIFP-08-32}

\maketitle

\section{Introduction}

 The great challenge in string phenomenology is the construction
of realistic string models, which do not have additional chiral 
exotic particles at low energy and can stabilize the moduli 
fields. Employing renormalization group equations, we may test 
such models at the upcoming Large Hadron Collider (LHC).  
In particular, the intersecting D-brane models on Type II 
orientifolds~\cite{JPEW}, where the chiral fermions arise 
from the intersections of D-branes in the
internal space~\cite{bdl} and the T-dual description in terms of
magnetized D-branes~\cite{bachas}, have been very
interesting during the last a few years~\cite{Blumenhagen:2005mu}.

In the beginning~\cite{Blumenhagen:2000wh}, a lot of 
non-supersymmetric three-family Standard-like models and 
Grand Unified Theories (GUTs) were constructed on Type IIA 
orientifolds with intersecting D6-branes.
However, these models generically 
have uncancelled Neveu-Schwarz-Neveu-Schwarz (NSNS) tadpoles
 and the gauge hierarchy problem. Later, 
semi-realistic supersymmetric Standard-like models and 
GUT models have been constructed in Type IIA theory on the
$\mathbf{T^6/(\Z_2\times \Z_2)}$ 
orientifold~\cite{CSU,CLL,Chen:2006sd} 
and other backgrounds~\cite{ListSUSYOthers}. We emphasize that 
only Pati-Salam like models can realize all the Yukawa 
couplings at the stringy tree level. Moreover, 
Pati-Salam like models have been 
constructed systematically in Type IIA theory on the 
$\mathbf{T^6/(\Z_2\times \Z_2)}$  
orientifold~\cite{CLL,Chen:2006sd}. 
Although the Standard Model (SM) fermion masses 
and mixings can be generated 
in one of these models~\cite{Chen:2007px},  we can not
stabilize the moduli fields and might not decouple all 
the chiral exotic particles. To stabilize the moduli via 
supergravity fluxes, the flux models on Type II orientifolds 
have also been constructed~\cite{Cascales:2003zp, 
Blumenhagen:2003vr, Marchesano:2004yq, Marchesano:2004xz,
Cvetic:2004xx, Cvetic:2005bn, Villadoro:2005cu, 
Camara:2005dc, Chen:2005cf, Chen:2006gd}. Especially, 
for the supersymmetric 
AdS vacua on Type IIA orientifolds with flux compactifications, 
the Ramond-Ramond (RR) tadpole cancellation 
conditions can be relaxed~\cite{Camara:2005dc, Chen:2006gd}. And 
then we can construct flux models that can explain the
SM fermion masses and mixings~\cite{Chen:2006gd}.
However, these models are in the AdS vacua and have
quite a few chiral exotic particles that are difficult 
to be decoupled. Recently, on  the Type IIB 
toroidal orientifold with the RR, NSNS, 
non-geometric and S-dual flux 
compactifications~\cite{Shelton:2005cf, Aldazabal:2006up,
Villadoro:2006ia}, we showed that  the RR tadpole 
cancellation conditions can be relaxed elegantly 
in the supersymmetric Minkowski vacua~\cite{Chen:2007af}.
Unfortunately, the Freed-Witten anomaly~\cite{Freed:1999vc} 
can give strong constraints on model building~\cite{Chen:2007af}, 
and the model in Ref.~\cite{Chen:2007af} indeed has the 
Freed-Witten anomaly that might be cancelled by introducing 
additional D-branes~\cite{Cascales:2003zp}.

In this paper, we revisit the Type IIB toroidal
orientifold with the RR, NSNS, non-geometric and S-dual 
flux compactifications~\cite{Aldazabal:2006up}. 
We present supersymmetric Minkowski vacua
where the RR tadpole cancellation conditions can be
relaxed and the Freed-Witten anomaly free conditions can be 
satisfied elegantly. We construct a realistic 
Pati-Salam like flux model without the Freed-Witten anomaly.
At the string scale, the gauge symmetry 
can be broken down to the SM gauge symmetry, the 
gauge coupling unification can be achieved naturally, 
and all the extra chiral exotic particles can be decoupled 
so that we obtain the supersymmetric SMs with/without 
SM singlet(s) below the string scale. The observed 
SM fermion masses and mixings can also be generated
since all the SM fermions and Higgs fields arise
from the intersections on the same two-torus. 
Moreover, the unified gauge coupling and the real
parts of the dilaton and K\"ahler moduli are
functions of the four-dimensional dilaton. And
the complex structure moduli and one linear 
combination of the imaginary parts of the K\"ahler 
moduli can be determined as functions of the 
fluxes and the dilaton. 
The systematical model building and the detailed
phenomenological discussions will be given 
elsewhere~\cite{CLN-L}.

This paper is organized as follows: in Section II we review
the Type IIB  model building and study the supersymmetric 
Minkowski flux vacua. We construct a realistic 
Pati-Salam like flux model and discuss its phenomenological
consequences in Section III. Discussion and conclusions are
presented in Section IV.

\section{Type IIB Flux Model Building}

Let us consider the Type IIB 
string theory compactified on a $\mathbf{T^6}$
orientifold where $\mathbf{T^{6}}$ is a six-torus
factorized as $\mathbf{T^{6}} = \mathbf{T^2} \times \mathbf{T^2}
\times \mathbf{T^2}$ whose complex coordinates are $z_i$, $i=1,\;
2,\; 3$ for the $i^{th}$ two-torus, respectively.
The orientifold projection is implemented by gauging the symmetry
$\Omega R$, where $\Omega$ is world-sheet parity, and $R$ is given
by
\begin{eqnarray}
R: (z_1,z_2,z_3) \to (-z_1, -z_2, -z_3)~.~\,    \label{orientifold}
\end{eqnarray}
Thus, the model contains 64 O3-planes. 
In order to cancel the negative RR charges from these
O3-planes, we introduce the magnetized  
D(3+2n)-branes which are filling up the 
four-dimensional Minkowski space-time and wrapping
2n-cycles on the compact manifold. Concretely, for one stack
of $N_a$ D-branes wrapped $m_a^i$ times on the $i^{th}$ 
two-torus $\mathbf{T^2_i}$, we turn on $n_a^i$ units of magnetic fluxes 
$F^i_a$ for the center of mass $U(1)_a$ gauge factor on $\mathbf{T^2_i}$, 
such that
\begin{eqnarray}
m_a^i \, \frac 1{2\pi}\, \int_{T^2_{\,i}} F_a^i \, = \, n_a^i ~,~\,
\label{monopole}
\end{eqnarray}
where $m_a^i$ can be half integer for tilted two-torus.
Then, the D9-, D7-, D5- and D3-branes contain 0, 1, 2 and 3 vanishing 
$m_a^i$s, respectively. Introducing for the $i^{th}$ two-torus 
the even homology classes $[{\bf 0}_i]$ and $[{\bf T}^2_i]$ for 
the point and two-torus, respectively, the vectors of the RR 
charges of the $a^{th}$ stack of D-branes and its image are
\begin{eqnarray}
 && [{ \Pi}_a]\, =\, \prod_{i=1}^3\, ( n_a^i [{\bf 0}_i] + m_a^i [{\bf T}^2_i] ), 
 \nonumber\\&&
[{\Pi}_a']\, =\, \prod_{i=1}^3\, ( n_a^i [{\bf 0}_i]- m_a^i [{\bf T}^2_i] )~,~
\label{homology class for D-branes}
\end{eqnarray}
respectively.
The ``intersection numbers'' in Type IIA language, which determine 
the chiral massless spectrum, are
\begin{eqnarray}
I_{ab}&=&[\Pi_a]\cdot[\Pi_b]=\prod_{i=1}^3(n_a^im_b^i-n_b^im_a^i)~.~
\label{intersections}
\end{eqnarray}
Moreover, for a stack of $N$ D(2n+3)-branes whose homology class
on $\mathbf{T^{6}}$ is (not) invariant under $\Omega R$, we obtain 
a ($U(N)$) $USp(2N)$  gauge symmetry with three (adjoint) 
anti-symmetric chiral superfields due to the orbifold projection.  
The physical spectrum is presented in Table \ref{spectrum}.

\begin{table}[t]
\caption{General spectrum  for magnetized D-branes on the  Type IIB
${\mathbf{T^6}}$ orientifold. }
\renewcommand{\arraystretch}{1.25}
\begin{center}
\begin{tabular}{|c|c|}
\hline {\bf Sector} & {\bf Representation}
 \\
\hline\hline
$aa$   & $U(N_a)$ vector multiplet  \\
       & 3 adjoint multiplets  \\
\hline
$ab+ba$   & $I_{ab}$ $(N_a,{\overline{N}}_b)$ multiplets  \\
\hline
$ab'+b'a$ & $I_{ab'}$ $(N_a,N_b)$ multiplets \\
\hline $aa'+a'a$ &$\frac 12 (I_{aa'} -  I_{aO3})\;\;$
symmetric multiplets \\
          & $\frac 12 (I_{aa'} +  I_{aO3}) \;\;$ 
anti-symmetric multiplets \\
\hline
\end{tabular}
\end{center}
\label{spectrum}
\end{table}

The flux models on Type IIB orientifolds with 
four-dimensional $N=1$ supersymmetry  are primarily
constrained by the  RR tadpole cancellation conditions that
will be given later, the four-dimensional $N=1$ supersymmetric
D-brane configurations, and the K-theory anomaly free conditions.
For D-branes with world-volume magnetic field
$F_a^i={n_a^i}/({m_a^i\chi_i})$ where $\chi_i$ is the area of 
 the $i^{th}$ two-torus $\mathbf{T^2_i}$
in string units,  the condition for the four-dimensional 
$N=1$ supersymmetric D-brane configurations is 
\begin{eqnarray}
\sum_i \left(\tan^{-1} (F_a^i)^{-1} + 
{\theta (n_a^i)} \pi \right)=0 ~~~{\rm mod}~ 2\pi~,~\,
\end{eqnarray}
where ${\theta (n_a^i)}=1$ for $n_a^i < 0$ and  
${\theta (n_a^i)}=0$ for $n_a^i \geq 0$.
The K-theory anomaly free conditions are~\cite{Marchesano:2004xz}
\begin{eqnarray}
 \sum_a N_a m_a^1 m_a^2 m_a^3 =  \sum_a N_a m_a^1 n_a^2 n_a^3
= \sum_a N_a n_a^1 m_a^2 n_a^3 
= \sum_a N_a n_a^1 n_a^2 m_a^3
=0 ~~~{\rm mod}~ 2~.~\,
\end{eqnarray}

We turn on the NSNS fluxes $h_0$ and $a_i$, 
the RR fluxes $e_i$ and $q_i$,
the non-geometric fluxes  $b_{ii}$, and the
S-dual fluxes $f_i$~\cite{Shelton:2005cf, Aldazabal:2006up,
Villadoro:2006ia, CLN-L}. To avoid subtleties,
these fluxes should be even integers due to the Dirac quantization.
For simplicity, we assume
\begin{eqnarray}
 a_i ~\equiv~ a ~,~~  e_i~ \equiv ~ e~,~~q_i ~\equiv~ q~,~~
{ b}_{ii} ~\equiv~ { \beta}_i~.~~\,
\end{eqnarray}
We can show that the constraints on fluxes from the Bianchi 
indentities are satisfied. The constraints on fluxes from 
the $SL(2, \mathbf{Z})$ S-duality invariance give
\begin{eqnarray}
a \beta_i ~=~ q f_i ~.~\,
\end{eqnarray}
The RR tadpole cancellation conditions are
\begin{eqnarray}
&& \sum_a N_an_a^1 n_a^2 n_a^3 =16 -{3\over 2} aq~,~
\nonumber\\&&
\sum_a N_a n_a^i m_a^j m_a^k = {1\over 2} q {\beta}_i~,~
\nonumber\\&&
N_{{\rm NS7}_i}=0~,~~N_{{\rm I7}_i}=0~,~\,
\end{eqnarray}
where $i\not= j \not= k \not= i$, and the $N_{{\rm NS7}_i}$
and $N_{{\rm I7}_i}$ denote the NS 7-brane charge and
the other 7-brane charge,
respectively~\cite{Aldazabal:2006up, CLN-L}.
Thus, if $aq <0$ and $q\beta_i < 0$, the RR tadpole cancellation
conditions are relaxed elegantly because $-aq/2$ and 
$-q \beta_i/2 $ only need to be even integers.
Moreover, we have seven moduli fields in the supergravity
theory basis: the dilaton $s$, three K\"ahler moduli
$t_i$, and three complex structure moduli
$u_i$. With the above fluxes, we can assume
\begin{eqnarray}
u_1=u_2=u_3 \equiv u~.~\,
\end{eqnarray}
Then the superpotential becomes
\begin{eqnarray}
{\cal W}=3 i e u -3 q u^2+s(ih_0-3au)-\beta_i t_i u -f_i s t_i~.~\,
\label{Flux-Superpotential}
\end{eqnarray}
In addition, the holomorphic gauge kinetic function for a generic stack of
D(2n+3)-branes  is given by~\cite{CLN-L, Cremades:2002te, Lust:2004cx}
\begin{eqnarray}
f_a &=& {1\over {\kappa_a}}\left(  n_a^1\,n_a^2\,n_a^3\,s-
n_a^1\,m_a^2\,m_a^3\,t_1 
-n_a^2\,m_a^1\,m_a^3\,t_2 -n_a^3\,m_a^1\,m_a^2\,t_3\right)~,~\,
\label{EQ-GKF}
\end{eqnarray}
where $\kappa_a$ is equal to  1 and 2 for $U(n)$ and $USp(2n)$,
respectively. And the K\"ahler potential for these moduli is of the usual
no-scale form~\cite{Lahanas:1986uc}
\begin{eqnarray}
{\cal K} = -{\rm ln}(s+{\bar s})-\sum_{i=1}^3 {\rm ln} (t_i + {\bar t}_i)
-\sum_{i=1}^3 {\rm ln} (u_i + {\bar u}_i)~.~\,
\end{eqnarray}
For the supersymmetric Minkowski vacua, we have
\begin{eqnarray}
{\cal W}={\partial_s  {\cal W}} ={\partial_{t_i}  {\cal W}}
={\partial_u  {\cal W}} =0 ~.~\,
\end{eqnarray}
From ${\partial_s  {\cal W}}={\partial_{t_i}  {\cal W}}=0$,
we obtain
\begin{eqnarray}
f_i t_i= ih_0-3au~,~~~ s=-{{q}\over a} u~,~\,
\label{Moduli-Relations}
\end{eqnarray}
 then the superpotential turns out to be 
\begin{eqnarray}
{\cal W} = \left( 3e -{{q h_0 }\over a}\right) i u~.~\,
\end{eqnarray}
Therefore, to satisfy ${\cal W}={\partial_u  {\cal W}} =0$,
we obtain
\begin{eqnarray}
3 e a ~=~ q h_0~.~\,
\end{eqnarray}
Because ${\rm Re} s > 0$,  ${\rm Re} t_i > 0$ and ${\rm Re} u_i > 0$,
we require
\begin{eqnarray}
{{f_i {\rm Re}t_i}\over a} < 0~,~~~{{q}\over a} < 0~.~\,
\end{eqnarray}
In short, in our constructions, we have fixed a linear combination
of  the K\"ahler moduli $t_i$ and 
the complex structure moduli $u$ 
as follows from Eq. (\ref{Moduli-Relations})
\begin{eqnarray}
&& f_i {\rm Re} t_i ~=~{{3a^2}\over q} {\rm Re} s~,~~
 {\rm Re} u ~=~-{{a}\over q}  {\rm Re} s~,~
\nonumber\\&&
f_i {\rm Im} t_i ~=~h_0+{{3a^2}\over q} {\rm Im} s~,~~
 {\rm Im} u ~=~-{{a}\over q}  {\rm Im} s~.~\,
\label{Mod-Relat-Comp}
\end{eqnarray}

In general, this kind of D-brane models might have the
Freed-Witten 
anomaly~\cite{Cascales:2003zp, Camara:2005dc, Freed:1999vc}.
In the world-volume of a generic stack of D-branes we have
a $U(1)$ gauge field whose scalar partner parametrizes the
D-brane position in compact space. These $U(1)$'s usually obtain
St\"uckelberg masses by swallowing RR scalar fields and then
decouple from the low-energy spectra.
In the mean time these scalars participate in the cancellation of
$U(1)$ gauge anomalies through a generalized Green-Schwarz 
mechanism~\cite{Ibanez:1998qp}. For the generic $a$ stack of 
D-branes, the $U(1)_a$ gauge field
couples to the RR fields in four dimensions as follows
\begin{eqnarray}
F^a \ \wedge \ N_a\sum_{I=0}^3 \ c_I^a C_I^{(2)} ~,~\,
\label{bf}
\end{eqnarray}
where $I=0,~1,~2,~3$, and
\begin{eqnarray}
 c_0^a \equiv m_a^1m_a^2m_a^3 \ ;\ c_1^a  \equiv m_a^1n_a^2n_a^3 \ ;\
c_2^a  \equiv n_a^1m_a^2n_a^3 \ ;\ c_3^a  \equiv n_a^1n_a^2m_a^3 \ \ ~,~\,
\end{eqnarray}
where  $C_0^{(2)}$ and $C_i^{(2)}$ are two-form fields that are 
Poincare duals to the ${\rm Im} s$ and ${\rm Im}  t_i$  fields 
in four dimensions, respectively. In terms of them the couplings
have a Higgs-like form
\begin{eqnarray}
 A_\mu ^a \partial ^\mu
( c_0^a\ {\rm Im} s- c_1^a {\rm Im}  t_1 - c_2^a {\rm Im}  t_2 -
c_3^a {\rm Im}  t_3) \ .
\label{higgs}
\end{eqnarray}
Thus,  certain linear combinations of the imaginary parts of the
$s$ and $t_i$ fields obtain masses by combining with open string
vector bosons living on the branes. In addition, these linear 
combinations of ${\rm Im} s$ and ${\rm Im}  t_i$ fields will
transform with a shift under $U(1)_a$ gauge transformations, like
Goldstone bosons do. If the shift symmetry for any D-brane stack
is violated by the flux potential, we shall have the Freed-Witten
anomaly~\cite{Freed:1999vc}. 
From the superpotential in Eq.~(\ref{Flux-Superpotential}),
we obtain that the flux potential may fix 
${\rm Im} s$ and one linear combination of ${\rm Im}  t_i$.
Thus, we derive the Freed-Witten anomaly free conditions
\begin{eqnarray}
c^a_0~=~0~,~~~ f_1 c^a_1+ f_2 c^a_2+ f_3 c^a_3~=~0~.~\,
\label{FW-I}
\end{eqnarray}
Or equivalently, we have
\begin{eqnarray}
c^a_0~=~0~,~~~ \beta_1 c^a_1+ \beta_2 c^a_2+ \beta_3 c^a_3~=~0~.~\,
\label{FW-II}
\end{eqnarray}

%%%%%%%%%%%%%%%%%%%%%%%%%%%%%%%%%%%%%%%%%%%%%%%%%%%%%%%%%%%%%%%%%%

%%%%%%%%%%%%%%%%%%%%%%%%%%%%%%%%%%%%%%%%%%%%%%%%%%%%%%%%%%%%%%%%%%

\section{A Realistic Model}

In this Section, we shall present a realistic model.
We choose the following fluxes
\begin{eqnarray}
&& a~=~8~,~ q~=~-2~,~\beta_1~=~2~,~\beta_2~=~6~,~\beta_3~=~6~,\,
\nonumber\\&&
f_1~=~-8~,~f_2~=~-24~,~f_3~=~-24~,~h_0~=~-12e~,~\,
\end{eqnarray}
where the flux $e$ is not fixed. We present the D-brane configurations 
and intersection numbers in Table~\ref{MI-Numbers}, and the 
resulting spectrum in Tables~\ref{Spectrum-I} and \ref{Spectrum-II}. 
One can easily check that our model satisfies the Freed-Witten anomaly 
free conditions in Eq. (\ref{FW-I}) or Eq. (\ref{FW-II}).

%%%%%%%%%%%%%%%%%%%%%%%%%%%%%%%%%%%%%%%%%%%%%%%%%%%%%%%%%%%%%%%%%%%%%%%%%%%%

%%%%%%%%%%%%%%%%%%%%%%%%%%%%%%%%%%%%%%%%%%%%%%%%%%%%%%%%%%%%%%%%%%%%%%%%%%%%

\begin{table}[h]
\begin{center}
\footnotesize
\renewcommand{\arraystretch}{.75}
\begin{tabular}{|c|c||ccc||c|c||c|c|c
||c|c|c|c|c|c|}
\hline

Stack & $N$ & ($n_1$,$l_1$) & ($n_2$,$l_2$) & ($n_3$,$l_3$) & A &
S & $b$  & $c$ & $c'$ & $d$ & $d'$ & $e$ & $e'$ & $f$ & $g$   \\
\hline \hline

$a$ & 4 & ( 2, 0) & ( 1,-1) & ( 1, 1) & 0(-1) & 0 & 3  & -3 &
0(-3) & -2 & 1 & 2 & -1 & 2 & -2  \\ \hline

$b$ & 2 & ( 1,-3) & ( 1, 1) & ( 2, 0) & 0(-3) & 0 & -  & 3 & 0(1) &
2 & -1 & 0 & 0 & 0 & 2     \\  \hline

$c$ & 2 & ( 1, 3) & ( 2, 0) & ( 1,-1) & 0(-3) & 0 & -  & - & - &
0 & 0 & -2 & 1 & -2 & 0  \\ \hline \hline

$d$ & 2 & ( 1, 1) & ( 2, 0) & ( 3,-1) & 0(-2) & 0(-1) & -  & - & - & -
& - & -1 & 0 & -2 & 0  \\ \hline

$e$ & 2 & ( 1,-1) & ( 3, 1) & ( 2, 0) & 0(-2) & 0(-1) & -  & - & - & -
& - & - & - & 0 & 2  \\ \hline

$f$ & 1 & ( 0, 2) & ( 0,-2) & ( 2, 0) & - & - & -  & - & - & -
& - & - & - & - & 0(-4)  \\ \hline

$g$ & 1 & ( 0, 2) & ( 2, 0) & ( 0,-2) & - & - & 
\multicolumn{9}{|c|}{$3\chi_1 = \chi_2= \chi_3$}
\\ \hline
\end{tabular}
\caption{D-brane configurations and intersection numbers
where $l_i\equiv 2m_i$. 
The complete gauge symmetry is $[U(4)_C \times U(2)_L \times
U(2)_R]_{\rm Observable}\times [U(2)^2 \times USp(2)^2]_{\rm Hidden}$, 
and the SM fermions and Higgs fields arise from the intersections
on the first two-torus. }
\label{MI-Numbers}
\end{center}
\end{table}

\begin{table}[h]
\begin{center}
\footnotesize
\renewcommand{\arraystretch}{.75}
\begin{tabular}{|c||c||c|c|c||c|c|c|}\hline
 & Quantum Number
& $Q_4$ & $Q_{2L}$ & $Q_{2R}$  & Field \\
\hline\hline
$ab$ & $3 \times (4,\bar{2},1,1,1,1,1)$ & 1 & -1 & 0  & $F_L(Q_L, L_L)$\\
$ac$ & $3\times (\bar{4},1,2,1,1,1,1)$ & -1 & 0 & $1$ & $F_R(Q_R, L_R)$\\
$bc$ & $3 \times (1,2,\bar{2},1,1,1,1)$ & 0 & 1 & -1   & 
$\Phi_i$($H_u^i$, $H_d^i$)\\
\hline\hline
$ac'$ & $3\times (4, 1, 2, 1, 1, 1, 1)$ & 1 & 0 & 1 & $X^i_{ac'}$ \\
  & $3\times ({\bar 4}, 1, {\bar 2}, 1, 1, 1, 1)$ & -1 & 0 & -1 
& $\overline{X}^i_{ac'}$ \\
\hline
$bc'$ & $1 \times (1,2, 2, 1, 1, 1, 1)$ & 0 & 1 & 1   & $\Phi'$($H'_u$, $H'_d$)\\
& $1 \times (1,\overline{2}, \overline{2}, 1, 1, 1, 1)$ & 0 & -1 & -1   & 
$\overline{\Phi}'$ \\
\hline
$aa'$ & $1\times (1, 1, 1, 1, 1, 1, 1)$ & 2 & 0 & 0 & $S_a$ \\
 & $1\times (\bar 1, 1, 1, 1, 1, 1, 1)$ & -2 & 0 & 0 & $\overline{S}_a$ \\
\hline
$ad$ & $2 \times (\bar 4,1,1,2,1,1,1)$ & -1 & 0 & 0  & $\overline{X}^i_{ad}$\\
$ad'$ & $1 \times ( 4,1,1,2,1,1,1)$ & 1 & 0 & 0  & $X_{ad'}$ \\
$ae$ & $2 \times (4,1,1,1,\bar 2,1,1)$ & 1 & 0 & 0  &  $X^i_{ae}$    \\
$ae'$ & $1 \times (\bar 4,1,1,1,\bar 2,1,1)$ & -1 & 0 & 0  
& $\overline{X}_{ae'}$ \\
$af$ & $2 \times (4,1,1,1,1,2,1)$ & 1 & 0 & 0  &   $X^i_{af}$   \\
$ag$ & $2 \times (\bar 4,1,1,1,1, 1, 2)$ & -1 & 0 & 0  
&  $\overline{X}^i_{ag}$\\
\hline
$bb'$ & $3\times (1, 1, 1, 1, 1, 1, 1)$ & 0 & 2 & 0 & $S^i_L$ \\
 & $3\times (1, \bar 1, 1, 1, 1, 1, 1)$ & 0 & -2 & 0 & $\overline{S}^i_L$ \\
\hline
$bd$ & $2 \times (1,2,1,\bar 2,1,1,1)$ & 0 & 1 & 0  &  $X^i_{bd}$\\
$bd'$ & $1 \times (1,\bar 2,1,\bar 2,1,1,1)$ & 0 & -1 & 0  
& $\overline{X}_{bd'}$\\
$bg$ & $2 \times (1,2,1,1,1,1, 2)$ & 0 & 1 & 0  & $X^i_{bg}$\\
\hline
$cc'$ & $3\times (1, 1, 1, 1, 1, 1, 1)$ & 0 & 0 & 2 & $S^i_R$ \\
 & $3\times (1, 1, \bar 1, 1, 1, 1, 1)$ & 0 & 0 & -2 & $\overline{S}^i_R$ \\
\hline
$ce$ & $2 \times (1,1,\bar 2,1,2,1,1)$ & 0 & 0 & -1  & $\overline{X}_{ce}^i$ \\
$ce'$ & $1 \times (1,1,2,1,2,1,1)$ & 0 & 0 & 1  &  $X_{ce'} $ \\
$cf$ & $2 \times (1,1,\bar 2,1,1,2,1)$ & 0 & 0 & -1  & $\overline{X}_{cf}^i$ \\
\hline
\end{tabular}
\caption{The chiral and vector-like superfields in the
observable sector, and their quantum
numbers under the gauge symmetry $U(4)_C\times U(2)_L\times U(2)_R
\times U(2)^2 \times USp(2)^2$.} 
\label{Spectrum-I}
\end{center}
\end{table}

%%%%%%%%%%%%%%%%%%%%%%%%%%%%%%%%%%%%%%%%%%%%%%%%%%%%%%%%%%%%%%%%%%%%%%%%%%%%

%%%%%%%%%%%%%%%%%%%%%%%%%%%%%%%%%%%%%%%%%%%%%%%%%%%%%%%%%%%%%%%%%%%%%%%%%%%%

\begin{table}[h]
\begin{center}
\footnotesize
\renewcommand{\arraystretch}{.75}
\begin{tabular}{|c||c||c|c|c||c|c|c|}\hline
 & Quantum Number
& $Q_4$ & $Q_{2L}$ & $Q_{2R}$  & Field \\
\hline\hline
$dd'$ & $1\times (1, 1, 1, 3, 1, 1, 1)$ & 0 & 0 & 0 & $T_d$ \\
& $1\times (1, 1, 1, \bar 3, 1, 1, 1)$ & 0 & 0 & 0 & $\overline{T}_d$ \\
& $2\times (1, 1, 1, 1, 1, 1, 1)$ & 0 & 0 & 0 & $S^i_d$ \\
 & $2\times (1, 1, 1, \bar 1, 1, 1, 1)$ & 0 & 0 &  & $\overline{S}^i_d$ \\
\hline
$de$ & $1 \times (1,1,1,\bar 2,2,1,1)$ & 0 & 0 & 0  & $\overline{X}_{de}$ \\
$df$ & $2 \times (1,1,1,\bar 2,1,2,1)$ & 0 & 0 & 0  & $\overline{X}_{df}^i$ \\
\hline
$ee'$ &  $1\times (1, 1, 1, 1, 3, 1, 1)$ & 0 & 0 & 0 & $T_e$ \\
& $1\times (1, 1, 1, 1, \bar 3, 1, 1)$ & 0 & 0 & 0 & $\overline{T}_e$ \\
& $2\times (1, 1, 1, 1, 1, 1, 1)$ & 0 & 0 & 0 & $S^i_e$ \\
 & $2\times (1, 1, 1,  1, \bar 1, 1, 1)$ & 0 & 0 &  & $\overline{S}^i_e$ \\
\hline
$eg$ & $2 \times (1,1,1,1,2,1,2)$ & 0 & 0 & 0  & $X^i_{eg}$  \\
\hline
$fg$ & $4\times (1, 1, 1, 1, 1, 2, 2)$ & 0 & 0 & 0 & $X^i_{fg}$ \\
 & $4\times (1, 1, 1,  1, 1, 2, 2)$ & 0 & 0 &  & $\overline{X}^i_{fg}$ \\
\hline
\end{tabular}
\caption{The chiral and vector-like superfields in the hidden sector, 
and their quantum
numbers under the gauge symmetry $U(4)_C\times U(2)_L\times U(2)_R
\times U(2)^2 \times USp(2)^2$.} 
\label{Spectrum-II}
\end{center}
\end{table}

%%%%%%%%%%%%%%%%%%%%%%%%%%%%%%%%%%%%%%%%%%%%%%%%%%%%%%%%%%%%%%%%%%%%%%%%%%%%

%%%%%%%%%%%%%%%%%%%%%%%%%%%%%%%%%%%%%%%%%%%%%%%%%%%%%%%%%%%%%%%%%%%%%%%%%%%%

In our model, the anomalies from the global $U(1)$s of the 
$U(4)_C$, $U(2)_L$, $U(2)_R$, $U(2)_d$ and $U(2)_e$ gauge symmetries
are cancelled by the generalized Green-Schwarz mechanism, 
and the gauge fields of 
the corresponding anomalous $U(1)$s obtain masses via the 
linear $B\wedge F$ couplings.  In addition, we can break
 the global $U(1)_a$, $U(1)_L$, $U(1)_R$, $U(1)_d$ and $U(1)_e$ 
gauge symmetries respectively of $U(4)_C$, $U(2)_L$, $U(2)_R$,
$U(2)_d$ and $U(2)_e$ by giving the string-scale vacuum 
expectation values (VEVs) to 
 $S_a$, $\overline{S}_a$,  $S^i_L$, $\overline{S}^i_L$, 
$S^i_R$, $\overline{S}^i_R$, $X_{de}$, $T_d$, $\overline{T}_d$,
$S^i_d$, $\overline{S}^i_d$,  $T_e$, $\overline{T}_e$,
$S^i_e$, and $\overline{S}^i_e$. Without loss of generality,
we can assume that their VEVs satisfy the D-flatness conditions for 
the global $U(1)_a$, $U(1)_L$, $U(1)_R$, $U(1)_d$ and $U(1)_e$ 
gauge symmetries. Thus, the
effective gauge symmetry in the observable sector
is indeed $SU(4)_C\times SU(2)_L\times SU(2)_R$.
In order to break the gauge symmetry down to
$ SU(3)_C\times SU(2)_L\times U(1)_{I_{3R}}\times U(1)_{B-L}$, on the 
first two-torus, we split the $a$ stack of
D-branes into $a_1$ and $a_2$ stacks with 3 and 1 D-branes,
respectively, and split the $c$ stack of D-branes into $c_1$ and
$c_2$ stacks with 1 D-brane for each one.
We can break the $U(1)_{I_{3R}}\times U(1)_{B-L}$ gauge
symmetry further down to the $U(1)_Y$ gauge symmetry by giving
the string-scale 
VEVs to the vector-like particles with quantum numbers
$({\bf { 1}, 1, 1/2, -1})$ and $({\bf { 1},
1, -1/2, 1})$ under $SU(3)_C\times SU(2)_L\times U(1)_{I_{3R}} \times
U(1)_{B-L} $ from $a_2 c_1'$ D-brane intersections. Similar to the
discussions in Ref.~\cite{Chen:2007px}, we can explain 
the SM fermion masses and mixings via 
the Higgs fields $H_u^i$, $H_u'$, $H_d^i$ and
$H_d'$ because all the SM fermions and Higgs fields arise 
from the intersections on the first two-torus. 
Note that we give the string-scale
VEVs to the fields $S^i_L$, $\overline{S}^i_L$, 
$S^i_R$, $\overline{S}^i_R$, $X_{de}$, $T_d$, $\overline{T}_d$,
$S^i_d$, $\overline{S}^i_d$,  $T_e$, $\overline{T}_e$,
$S^i_e$, and $\overline{S}^i_e$,  the chiral exotic particles 
can obtain masses around the string scale via the following 
superpotential from three-point and four-point functions
\begin{eqnarray}
W & \supset & (\overline{T}_d+\overline{S}_d^i) \overline{X}_{ad}^j
X_{ad'} + (T_e +S_e^i) X_{ae}^j \overline{X}_{ae'}
+X_{de}(\overline{X}_{ad}^i X_{ae}^j +  X_{ad'} \overline{X}_{ae'})
\nonumber\\&&
+(X_{fg}^i+\overline{X}_{fg}^i) X_{af}^j \overline{X}_{ag}^k
+(T_d +S_d^i) X_{bd}^j \overline{X}_{bd'}
+\overline{S}^i_L X_{bg}^j X_{bg}^k
\nonumber\\&&
+ (\overline{T}_e+\overline{S}_e^i) \overline{X}_{ce}^j
X_{ce'} + S_R^i \overline{X}_{cf}^j \overline{X}_{cf}^k
+ ({T}_d+{S}_d^i) \overline{X}_{df}^j
\overline{X}_{df}^k 
\nonumber\\&&
+  (\overline{T}_e+\overline{S}_e^i) X_{eg}^j X_{eg}^k
+ {1\over {M_{\rm St}}} \left( \overline{S}_L^i 
({T}_d+{S}_d^j) X_{bd}^k  X_{bd}^l 
\right.\nonumber\\&&\left.
+ S_R^i (\overline{T}_e+\overline{S}_e^j) \overline{X}_{ce}^k
\overline{X}_{ce}^l 
+ ({T}_d+{S}_d^i) (\overline{T}_e+\overline{S}_e^j) 
\overline{X}_{de} \overline{X}_{de} \right) ~,~\,
\end{eqnarray}
where $M_{\rm St}$ is the string scale, and
we neglect the ${\cal{O}}(1)$ coefficients in this paper. 
In addition, we can
decouple all the Higgs bidoublets close to the string scale  
except one pair of the linear combinations of the Higgs doublets 
for the electroweak symmetry breaking at the low
energy by fine-tuning the following superpotential
\begin{eqnarray}
W & \supset &  \Phi_i ( \overline{S}_L^j \Phi' + S_R^j \overline{\Phi}')
+{1\over {M_{\rm St}}} \left( \overline{S}_L^i S_R^j \Phi_k \Phi_l 
+\overline{S}_L^i \overline{S}_R^j \Phi' \Phi' 
+ S_L^i S_R^j \overline{\Phi}' \overline{\Phi}' \right)
~.~\,
\end{eqnarray}
In short, below the string scale, we have the supersymmetric
SMs which may have zero, one or a few SM singlets from $S_L^i$, 
$\overline{S}_L^i$, $S^i_R$, and  $\overline{S}^i_R$. Then the
upper bound on the lightest CP-even Higgs boson mass in the
minimal supersymmetric SM can
be relaxed if we have the SM singlet(s) at low energy~\cite{Li:2006xb}.

Next, we consider the gauge coupling unification and moduli stabilization.
Note that $3\chi_1=\chi_2=\chi_3$ as given in Table~\ref{MI-Numbers}
are derived from the supersymmetric D-brane configurations, we define 
\begin{eqnarray}
\chi_1~\equiv~\chi~,~ \chi_2=\chi_3\equiv 3 \chi~.~\,
\end{eqnarray}
Thus, the real parts of the dilaton and K\"ahler moduli 
in our model are~\cite{CLN-L}
\begin{eqnarray}
&&  {\rm Re} s = {{ e^{-\phi_4}}\over {6 \pi \chi {\sqrt \chi}}}~,~
{\rm Re} t_1 = {{3 {\sqrt \chi} e^{-\phi_4}}\over {2 \pi}}~,~
\nonumber\\&&
{\rm Re} t_2 ={{{\sqrt \chi} e^{-\phi_4}}\over {2 \pi}}~,~
{\rm Re} t_3 = {{{\sqrt \chi} e^{-\phi_4}}\over {2 \pi}}~,~\,
\label{Moduli}
\end{eqnarray}
where $\phi_4$ is the four-dimensional dilaton. From Eq. (\ref{EQ-GKF}),
 we obtain that the SM gauge couplings are unified at the string
scale as follows
\begin{eqnarray}
g_{SU(3)_C}^{-2} = g_{SU(2)_L}^{-2} ={3\over 5} g_{U(1)_Y}^{-2}
={{ e^{-\phi_4}}\over {2 \pi}} \left( {2\over {3\chi {\sqrt \chi}}}
+{{3 {\sqrt \chi}}\over 2} \right)~.~\,
\end{eqnarray}
From the real part of the first equation in Eq. (\ref{Moduli-Relations})
or the first equation in Eq. (\ref{Mod-Relat-Comp}),
we obtain
\begin{eqnarray}
\chi &=& {2\over 3}~.~
\end{eqnarray}
Therefore, $\chi_i$ are determined by the supersymmetric D-brane
configurations and the conditions for the supersymmetric 
Minkowski vacua.  
Using the unified gauge coupling $g^{2} \simeq 0.513$
in supersymmetric SMs, we get
\begin{eqnarray}
 \phi_4 \simeq - 1.61~.~\,
\end{eqnarray}

From Eq. (\ref{Mod-Relat-Comp}), we obtain
\begin{eqnarray}
&&  {\rm Re} u ~=~4  {\rm Re} s~,~ 
{\rm Im} u ~=~ 4  {\rm Im} s~,~
\nonumber\\&&
8 {\rm Im} t_1 + 24 {\rm Im} t_2 
+ 24 {\rm Im} t_3~=~-h_0+96 {\rm Im} s~.~\,
\end{eqnarray}

%%%%%%%%%%%%%%%%%%%%%%%%%%%%%%%%%%%%%%%%%%%%%%%%%%%%%%%%%%%%%%%%%%%

%%%%%%%%%%%%%%%%%%%%%%%%%%%%%%%%%%%%%%%%%%%%%%%%%%%%%%%%%%%%%%%%%%%

%%%%%%%%%%%%%%%%%%%%%%%%%%%%%%%%%%%%%%%%%%%%%%%%%%%%%%%%%%%%%%%%%%%

%%%%%%%%%%%%%%%%%%%%%%%%%%%%%%%%%%%%%%%%%%%%%%%%%%%%%%%%%%%%%%%%%%%

\section{Discussion and Conclusions}

We showed that the RR tadpole cancellation conditions can be 
relaxed and the Freed-Witten anomaly can be cancelled elegantly 
in the supersymmetric Minkowski vacua on the Type IIB toroidal 
orientifold with general flux compactifications. And we 
presented a realistic Pati-Salam like flux model in details. 
In this model, we can break 
the gauge symmetry down to the SM gauge symmetry,
realize the gauge coupling unification,
and decouple  all the extra chiral exotic particles
around the string scale. We can also generate the
observed SM fermion masses and mixings.
Futhermore,  the unified gauge coupling and the real
parts of the dilaton and K\"ahler moduli are
functions of the four-dimensional dilaton. And
the complex structure moduli and one linear 
combination of the imaginary parts of the K\"ahler 
moduli can be determined as functions of the 
fluxes and the dilaton.

\section*{Acknowledgments}

This research was supported in part by
the Mitchell-Heep Chair in High Energy Physics (CMC), by the
Cambridge-Mitchell Collaboration in Theoretical Cosmology 
and by the NSFC grant 10821504 (TL),
and by the DOE grant DE-FG03-95-Er-40917 (DVN).

%%%%%%%%%%%%%%%%%%%%%%%%%%%%%%%%%%%%%%%%%%%%%%%%%%%%%%%%%%%%%%%%%%%%%%%%%%%%


\begin{thebibliography}{99}



\bibitem{JPEW}

J.~Polchinski and E.~Witten, Nucl.\ Phys.\ B {\bf 460}, 525
(1996).



\bibitem{bdl}

M.~Berkooz, M.\,R.~Douglas and R.\,G.~Leigh, Nucl. Phys. B {\bf
480} (1996) 265.



\bibitem{bachas}

C.~Bachas, hep-th/9503030.

%%CITATION = HEP-TH 9503030;%%



%\cite{Blumenhagen:2005mu}

\bibitem{Blumenhagen:2005mu}

  R.~Blumenhagen, M.~Cveti\v c, P.~Langacker and G.~Shiu,
  %``Toward realistic intersecting D-brane models,''
  Ann.\ Rev.\ Nucl.\ Part.\ Sci.\  {\bf 55}, 71 (2005),
and the references therein.

%  [arXiv:hep-th/0502005].

  %%CITATION = ARNUA,55,71;%%



%\cite{Blumenhagen:2000wh}

\bibitem{Blumenhagen:2000wh}

  R.~Blumenhagen, L.~G\"orlich, B.~K\"ors and D.~L\"ust,
  %``Noncommutative compactifications of type I strings on tori with  magnetic
  %background flux,''
  JHEP {\bf 0010}, 006 (2000);
%  [arXiv:hep-th/0007024].
  %%CITATION = HEP-TH 0007024;%%
  %%Cited 208 times in SPIRES-HEP
  C.~Angelantonj, I.~Antoniadis, E.~Dudas and A.~Sagnotti,
  %``Type-I strings on magnetised orbifolds and brane transmutation,''
  Phys.\ Lett.\ B {\bf 489}, 223 (2000).
%  [arXiv:hep-th/0007090].
  %%CITATION = HEP-TH 0007090;%%
  %%Cited 135 times in SPIRES-HEP





\bibitem{CSU}

M.~Cveti\v c, G.~Shiu and A.\,M.~Uranga, Phys.\ Rev.\ Lett.\ {\bf
87}, 201801 (2001);
%%CITATION = HEP-TH 0107143;%%
Nucl.\ Phys.\ B {\bf 615}, 3 (2001).
%%CITATION = HEP-TH 0107166;%%







\bibitem{CLL}

M.~Cveti\v c, T.~Li and T.~Liu,
%``Supersymmetric Pati-Salam models from intersecting D6-branes: A road to the
%standard model,''
Nucl.\ Phys.\ B {\bf 698}, 163 (2004);
% hep-th/0403061.
%%CITATION = HEP-TH 0403061;%%
M.~Cvetic, P.~Langacker, T.~Li and T.~Liu,
  %``D6-brane splitting on type IIA orientifolds,''
  Nucl.\ Phys.\  B {\bf 709}, 241 (2005).
%  [arXiv:hep-th/0407178].
  %%CITATION = NUPHA,B709,241;%%


%\cite{Chen:2006sd}
\bibitem{Chen:2006sd}
  C.~M.~Chen, V.~E.~Mayes and D.~V.~Nanopoulos,
  %``MSSM via Pati-Salam from Intersecting Branes on $T^6/(\mathbf{\Z_2}
  %\times \mathbf{\Z_2'})$,''
  Phys.\ Lett.\  B {\bf 648}, 301 (2007).
%  [arXiv:hep-th/0612087].
  %%CITATION = PHLTA,B648,301;%%


\bibitem{ListSUSYOthers}

R.~Blumenhagen, L.~G\"orlich and T.~Ott, JHEP {\bf 0301}, 021
(2003);
% hep-th/0211059;
%%CITATION = HEP-TH 0211059;%%
G.~Honecker, Nucl.\ Phys.\  {\bf B666}, 175 (2003);
%  hep-th/0303015;
%%CITATION = HEP-TH 0303015;%%
G.~Honecker and T.~Ott,
%``Getting just the supersymmetric standard model at intersecting branes on the
%Z(6)-orientifold,''
Phys.\ Rev.\ D {\bf 70}, 126010 (2004) [Erratum-ibid.\ D {\bf 71},
069902 (2005)].
% hep-th/0404055.
%%CITATION = HEP-TH 0404055;%%


%\cite{Chen:2007px}
\bibitem{Chen:2007px}
  C.~M.~Chen, T.~Li, V.~E.~Mayes and D.~V.~Nanopoulos,
  %``A Realistic World from Intersecting D6-Branes,''
  Phys.\ Lett.\  B {\bf 665}, 267 (2008);
%  [arXiv:hep-th/0703280].
  %%CITATION = PHLTA,B665,267;%%
%``Towards realistic supersymmetric spectra and Yukawa textures from
  %intersecti ng branes,''
  Phys.\ Rev.\  D {\bf 77}, 125023 (2008).
%  [arXiv:0711.0396 [hep-ph]].
  %%CITATION = PHRVA,D77,125023;%%


%%%%%%%%%%%%%%%%%%%%%%%%%%%%%%%%%%%%%%%%%%%%%%%%%%%%%%%%


%\cite{Cascales:2003zp}
\bibitem{Cascales:2003zp}
  J.~F.~G.~Cascales and A.~M.~Uranga,
  %``Chiral 4d N = 1 string vacua with D-branes and NSNS and RR fluxes,''
  JHEP {\bf 0305}, 011 (2003).
%  [arXiv:hep-th/0303024].
  %%CITATION = JHEPA,0305,011;%%


%\cite{Blumenhagen:2003vr}
\bibitem{Blumenhagen:2003vr}
  R.~Blumenhagen, D.~Lust and T.~R.~Taylor,
  %``Moduli stabilization in chiral type IIB orientifold models with fluxes,''
  Nucl.\ Phys.\  B {\bf 663}, 319 (2003).
%  [arXiv:hep-th/0303016].
  %%CITATION = NUPHA,B663,319;%%


%%%%%%%%%%%%%%%%%%%%%%%%%%%%%%%%%%%%%%%%%%%%%%%%%%%%%%%%

%\cite{Marchesano:2004yq}
\bibitem{Marchesano:2004yq}
  F.~Marchesano and G.~Shiu,
  %``MSSM vacua from flux compactifications,''
  Phys.\ Rev.\  D {\bf 71}, 011701 (2005).
%  [arXiv:hep-th/0408059].
  %%CITATION = PHRVA,D71,011701;%%

%\cite{Marchesano:2004xz}
\bibitem{Marchesano:2004xz}
  F.~Marchesano and G.~Shiu,
  %``Building MSSM flux vacua,''
  JHEP {\bf 0411}, 041 (2004).
%  [arXiv:hep-th/0409132].
  %%CITATION = JHEPA,0411,041;%%


%\cite{Cvetic:2004xx}
\bibitem{Cvetic:2004xx}
  M.~Cvetic and T.~Liu,
  %``Three-family supersymmetric standard models, flux compactification and
  %moduli stabilization,''
  Phys.\ Lett.\  B {\bf 610}, 122 (2005).
%  [arXiv:hep-th/0409032].
  %%CITATION = PHLTA,B610,122;%%

%\cite{Cvetic:2005bn}
\bibitem{Cvetic:2005bn}
  M.~Cvetic, T.~Li and T.~Liu,
  %``Standard-like Models as Type IIB Flux Vacua,''
  Phys.\ Rev.\  D {\bf 71}, 106008 (2005).
%  [arXiv:hep-th/0501041].
  %%CITATION = PHRVA,D71,106008;%%


%%%%%%%%%%%%%%%%%%%%%%%%%%%%%%%%%%%%%%%%%%%%%%%%%%%%%%%%




%\cite{Villadoro:2005cu}
\bibitem{Villadoro:2005cu}
  G.~Villadoro and F.~Zwirner,
  %``N = 1 effective potential from dual type-IIA D6/O6 orientifolds with
  %general fluxes,''
  JHEP {\bf 0506}, 047 (2005).
%  [arXiv:hep-th/0503169].
  %%CITATION = JHEPA,0506,047;%%


%\cite{Camara:2005dc}
\bibitem{Camara:2005dc}
  P.~G.~Camara, A.~Font and L.~E.~Ibanez,
  %``Fluxes, moduli fixing and MSSM-like vacua in a simple IIA orientifold,''
  JHEP {\bf 0509}, 013 (2005).
%  [arXiv:hep-th/0506066].
  %%CITATION = JHEPA,0509,013;%%


%%%%%%%%%%%%%%%%%%%%%%%%%%%%%%%%%%%%%%%%%%%%%%%%%%%%%%%%

%\cite{Chen:2005cf}
\bibitem{Chen:2005cf}
  C.~M.~Chen, V.~E.~Mayes and D.~V.~Nanopoulos,
  %``Flipped SU(5) from D-branes with type IIB fluxes,''
  Phys.\ Lett.\  B {\bf 633}, 618 (2006).
%  [arXiv:hep-th/0511135].
  %%CITATION = PHLTA,B633,618;%%


%\cite{Chen:2006gd}

\bibitem{Chen:2006gd}  
C.-M.~Chen, T.~Li and D.\,V.~Nanopoulos,
  %``Type IIA Pati-Salam flux vacua,''
  Nucl.\ Phys.\  B {\bf 740}, 79 (2006).
%  [arXiv:hep-th/0601064].
  %%CITATION = NUPHA,B740,79;%%




%\cite{Shelton:2005cf}
\bibitem{Shelton:2005cf}
  J.~Shelton, W.~Taylor and B.~Wecht,
  %``Nongeometric Flux Compactifications,''
  JHEP {\bf 0510}, 085 (2005).
%  [arXiv:hep-th/0508133].
  %%CITATION = JHEPA,0510,085;%%


%\cite{Aldazabal:2006up}
\bibitem{Aldazabal:2006up}
  G.~Aldazabal, P.~G.~Camara, A.~Font and L.~E.~Ibanez,
  %``More dual fluxes and moduli fixing,''
  JHEP {\bf 0605}, 070 (2006).
%  [arXiv:hep-th/0602089].
  %%CITATION = JHEPA,0605,070;%%


%\cite{Villadoro:2006ia}
\bibitem{Villadoro:2006ia}
  G.~Villadoro and F.~Zwirner,
  %``D terms from D-branes, gauge invariance and moduli stabilization in  flux
  %compactifications,''
  JHEP {\bf 0603}, 087 (2006).
%  [arXiv:hep-th/0602120].
  %%CITATION = JHEPA,0603,087;%%


%\cite{Chen:2007af}
\bibitem{Chen:2007af}
  C.~M.~Chen, T.~Li, Y.~Liu and D.~V.~Nanopoulos,
  %``Realistic Type IIB Supersymmetric Minkowski Flux Vacua,''
  Phys.\ Lett.\  B {\bf 668}, 63 (2008).
%  [arXiv:0711.2679 [hep-th]].
  %%CITATION = PHLTA,B668,63;%%


%\cite{Freed:1999vc}
\bibitem{Freed:1999vc}
  D.~S.~Freed and E.~Witten,
  %``Anomalies in string theory with D-branes,''
  arXiv:hep-th/9907189.
  %%CITATION = HEP-TH/9907189;%%


%\cite{CLN-L}
\bibitem{CLN-L}
C.-M.~Chen, T.~Li, and D.~V.~Nanopoulos,
in preparation.


%\cite{Cremades:2002te}
\bibitem{Cremades:2002te}
  D.~Cremades, L.~E.~Ibanez and F.~Marchesano,
  %``SUSY quivers, intersecting branes and the modest hierarchy problem,''
  JHEP {\bf 0207}, 009 (2002).
%  [arXiv:hep-th/0201205].
  %%CITATION = JHEPA,0207,009;%%


%\cite{Lust:2004cx}
\bibitem{Lust:2004cx}
  D.~Lust, P.~Mayr, R.~Richter and S.~Stieberger,
  %``Scattering of gauge, matter, and moduli fields from intersecting  branes,''
  Nucl.\ Phys.\  B {\bf 696}, 205 (2004).
%  [arXiv:hep-th/0404134].
  %%CITATION = NUPHA,B696,205;%%


%\cite{Lahanas:1986uc}
\bibitem{Lahanas:1986uc}
  A.~B.~Lahanas and D.~V.~Nanopoulos,
  %``The Road to No Scale Supergravity,''
  Phys.\ Rept.\  {\bf 145}, 1 (1987).
  %%CITATION = PRPLC,145,1;%%


%\cite{Ibanez:1998qp}
\bibitem{Ibanez:1998qp}
  L.~E.~Ibanez, R.~Rabadan and A.~M.~Uranga,
  %``Anomalous U(1)'s in type I and type IIB D = 4, N = 1 string vacua,''
  Nucl.\ Phys.\  B {\bf 542}, 112 (1999).
%  [arXiv:hep-th/9808139].
  %%CITATION = NUPHA,B542,112;%%


%\cite{Li:2006xb}
\bibitem{Li:2006xb}
  T.~Li,
  %``String inspired singlet extensions of the minimal supersymmetric  standard
  %model,''
  Phys.\ Lett.\  B {\bf 653}, 338 (2007).
%  [arXiv:hep-ph/0612359].
  %%CITATION = PHLTA,B653,338;%%


\end{thebibliography}
\end{document}